\documentclass[a4paper,11pt]{article}
\usepackage{jheppub} 
\usepackage{lineno}
\usepackage{makecell}
\usepackage{float}
\newcommand{\tr}{\operatorname{tr}}
\newcommand{\K}{\mathcal{K}}
\newcommand{\BH}{\mathcal{B}(\mathcal{H})}
\newcommand{\mL}{\mathcal{L}}
\newcommand{\E}{\mathbb{E}}

\title{\boldmath Operator Krylov complexity in random matrix theory}

\author[a]{Haifeng Tang}
\affiliation[a]{Department of Physics, Stanford University, Stanford, California 94305, USA}
\emailAdd{hftang@stanford.edu}

\abstract{
Krylov complexity, as a novel measure of operator complexity under Heisenberg evolution, exhibits many interesting universal behaviors and also bounds many other complexity measures. In this work, we study Krylov complexity $\K(t)$ in Random Matrix Theory (RMT). In large $N$ limit: (1) For infinite temperature, we analytically show that the Lanczos coefficient $\{b_n\}$ saturate to constant plateau $\lim\limits_{n\rightarrow\infty}b_n=b$, rendering a linear growing complexity $\K(t)\sim t$, in contrast to the exponential-in-time growth in chaotic local systems in thermodynamic limit. After numerically comparing this plateau value $b$ to a large class of chaotic local quantum systems, we find that up to small fluctuations, it actually bounds the $\{b_n\}$ in chaotic local quantum systems. Therefore we conjecture that in chaotic local quantum systems after scrambling time, the speed of linear growth of Krylov complexity cannot be larger than that in RMT. (2) For low temperature, we analytically show that $b_n$ will first exhibit linear growth with $n$, whose slope saturates the famous chaos bound. After hitting the same plateau $b$, $b_n$ will then remain constant. This indicates $\K(t)\sim e^{2\pi t/\beta}$ before scrambling time $t_*\sim O(\beta\log\beta)$, and after that it will grow linearly in time, with the same speed as in infinite temperature. We finally remark on the effect of finite $N$ corrections.}


\begin{document}
\maketitle
\flushbottom

\section{Introduction}
\label{sec:introduction}
Operator complexity describes the phenomenon that a simple operator $O$ becomes complex under Heisenberg evolution $O(t)$ in chaotic local quantum systems. Many complexity measures~\cite{Roberts2017,Cotler2017chaos,Jefferson2017,Roberts2018,Yang2018,Khan2018,Qi2019,Lucas2019,Balasubramanian2020,Balasubramanian2021} make this manifest. An intuitive one is the operator size and OTOC~\cite{Roberts2018,Qi2019,Lucas2019}, which counts the average size of spatial support for an operator written in the basis made from tensor product of simple operators. Since operator size is bounded by the size of the whole system, then one would expect that size would cease to grow and saturate soon after it spreads onto the whole system at the time scale of scrambling time $t_*$~\cite{Maldacena2016bound}. However, since the Heisenberg evolution proceeds, one would expect that the operator is still growing more and more complex, simply not in the sense of the size of its spatial support. 

\paragraph{Behaviour of Krylov complexity in generic chaotic local systems.} Krylov complexity~\cite{Parker2019}, denoted as $\K(t)$, proposed as a measure of operator complexity, achieves this goal. Namely, it successfully captures the entire time scale and different stages of complexity dynamics. In generic chaotic local quantum systems and starting from a simple operator, the dynamics of $\K(t)$ often exhibit three stages:
\begin{itemize}
\item[1.] \textit{Stage one: exponential-in-time growth.} $\K(t)\sim e^{2\alpha t}$. This exponential growth will extend up to scrambling time $t_*\sim\log(S)$ and $\K(t)$ reaches the value of order $O(S)$, where $S$ is the number of degree of freedom. The complexity growth in this stage is mainly due to spreading a simple operator to the whole system, therefore describing the same physics as in operator size. The temperature-dependent exponential coefficient $\alpha$ is related and actually bounds the Lyapunov exponential of OTOC~\cite{Maldecena2016,Maldacena2016bound}.

\item[2.]\textit{Stage two: linear-in-time growth.} $\K(t)\sim v t$. Up to scrambling time in stage one, the growth of `complexity in real space' has finished. In this linear growth region, $\K(t)$ probes the growth of `complexity in Hilbert space'. Roughly speaking, $\K(t)$ in this stage characterizes the fact that $O(t)$ explores more and more linearly independent basis that it has not been occupied before in the operator vector space as large as $O(e^{2S})$. This stage will last up to a time scale $O(e^{2S})$ where $\K(t)$ reaches the value of order $O(e^{2S})$~\cite{Rabinovici2021}, the dimension of operator vector space, as claimed. 
\item[3.] \textit{Stage three: remain constant}. $\K(t)\sim \text{Const}$. At the end of linear-in-time stage, $O(t)$ has explored the edge of Krylov subspace and occupied as much linearly independent operator basis as it can, therefore it will remain constant afterward. This constant value is precisely $\frac{1}{2}K$~\cite{Rabinovici2021}, which is the half of the Krylov subspace dimension $K\sim O(e^{2S})$. The prefactor $\frac{1}{2}$ means that $O(t)$ is evenly distributed on every basis of Krylov subspace, indicating that it is fully scrambled. There might be possible Poincar\'e recurrence at the time of order $O(e^{e^{2S}})$ corresponding to some doubly-non-perturbative physics that is beyond our scope~\cite{Rabinovici2021}.
\end{itemize}

\begin{figure}[t]
    \centering
    \includegraphics[width=0.9\textwidth]{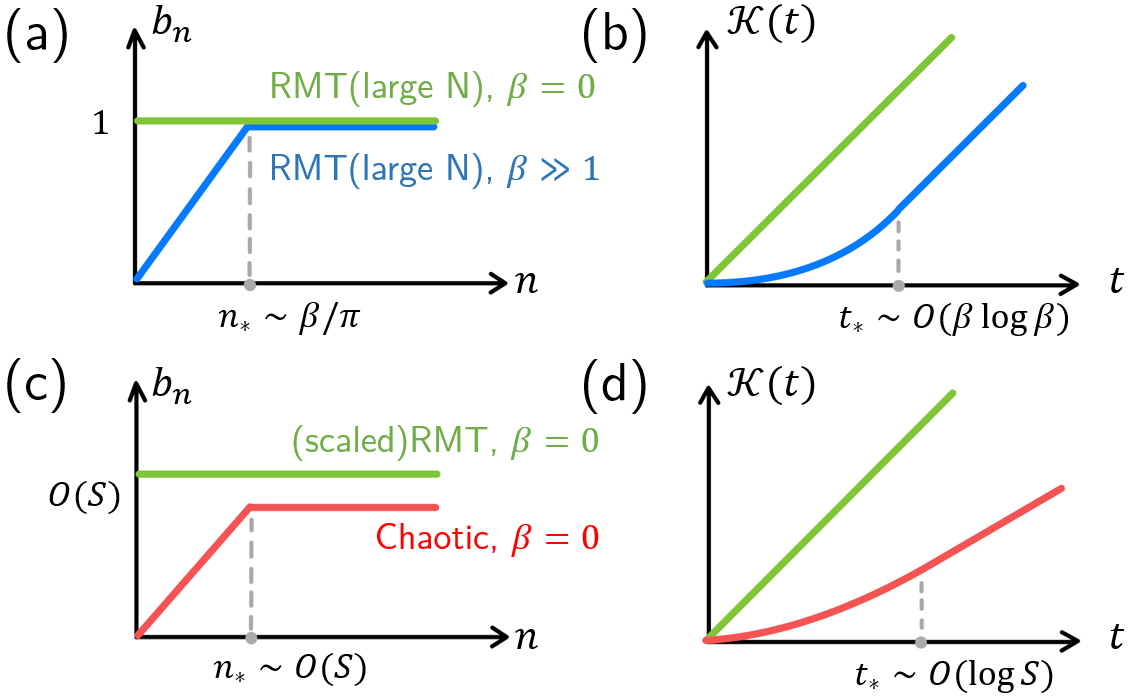}
    \caption{Schematic plot of the main result of this work. In \textbf{(a)} and \textbf{(b)}, we show the behavior of Krylov complexity $\K(t)$ and Lanczos coefficient $\{b_n\}$ in Random Matrix Theory in large $N$ limit with finite or infinite temperature. The plateau value equating unity comes from normalization of radius of Wigner semicircle to be unity.
    In \textbf{(c)} and \textbf{(d)}, we compare the random matrix results with generic chaotic local quantum systems.}
    \label{fig:schematic}
\end{figure}

\paragraph{Motivation of our work. } In this paper, we mainly focus on the linear-in-time stage, and sometime on exponential-in-time stage. Our motivation is that, at the conjunction of stage one and two, $O(t)$ has just finished its `scrambling in real space', and begun to start its `scrambling in Hilbert space'. Therefore in stage two, $O(t)$ completely abandoned any spatial locality structure, or equivalently, the local Hamiltonian $H$ no longer looks local from the perspective of $O(t)$. Therefore, in this stage, $H$ would act like a chaotic Hamiltonian without spatial locality, resembling a typical random matrix. 

This is our main reasoning and motivation for studying Krylov complexity in Random Matrix Theory (RMT), which intuitively should dominate the behavior of complexity growth after scrambling time. This is consistent with the fact that in stage one, the exponential growth in time crucially comes from the argument of the locality of $O(t)$ and $H$~\cite{Parker2019}. Without locality, we would expect the growth rate to decrease to linear. 

Another important motivation is that the linear growth of complexity characterizes the expanding volume of the Einstein-Rosen bridge in the dual holographic gravity~\cite{susskind2014computational}. We expect our RMT analysis would shed light on bulk gravitational systems.

\paragraph{Main results and structure of the paper.}
In this paper, we obtain several analytical results of Krylov complexity in a specific random matrix ensemble, called GUE (Gaussian Unitary Ensemble) in large $N$ limit, thanks to the large $N$ factorization. 

In section~\ref{sec:review on K complexity}, we review the construction of Krylov basis which originates from an orthogonalization procedure, the definition of Krylov complexity, and associated Lanczos coefficient $\{b_n\}$.

In section~\ref{sec:infinite twmperature}, we study the Krylov complexity of arbitrary traceless initial operator $O$, and we show that if we properly scale the radius of Wigner semicircle to unity, for Lanczos coefficient $\{b_n\}$, we have:
\begin{equation}
\lim\limits_{n\rightarrow\infty}\lim\limits_{N\rightarrow\infty}b_n=1
\end{equation}
This indicates that in large $N$ limit, $\K(t)$ in RMT grows linearly in time, in contrast to the chaotic local quantum systems in thermodynamic limit, whose K-complexity exhibits an exponential growth (in thermodynamic limit, scrambling time $t_*$ goes to infinity, therefore chaotic local system only have stage one, without stage two and three). We remark that this does not mean K-complexity in RMT is smaller than that in chaotic local systems. This is because, in RMT theory, we usually scale the Hamiltonian such that its eigenenergy is of order $O(1)$. But in chaotic local systems, the energy spectrum is extensive which is of order $O(S)$. Since the energy scale is conjugate to the time scale, we should properly rescale our RMT result.

Indeed, in section~\ref{sec:numerical compare chaotic}, we numerically compared the Lanczos coefficient $\{b_n\}$ in generic chaotic local systems in stage two with that from properly rescaled RMT, and find that the former is generally of the same order and smaller than the latter up to fluctuation. This indicates that the stage two growth in chaotic local systems originates the onset of random matrix behavior, as claimed in the motivation. Due to various empirical findings from numerics and theoretic reasoning, we therefore conjecture that stage two growth in chaotic local systems is bounded by scaled RMT growth. This is schematically summarized in figure~\ref{fig:schematic}(c), (d).

Nevertheless, the complexity dynamics of RMT is by itself interesting. In section~\ref{sec:finite temeprature}, still working in large $N$ limit, we deviate from infinite temperature and study K-complexity in RMT at finite and especially low temperature. We analytically show that $\{b_n\}$ will first grow linear in $n$ with slope $\alpha$ proportional to temperature $T$, until it reaches the plateau value $b_n=1$, the same as in the case of infinite temperature. Through numerical simulations, we fix $\alpha=\pi\beta^{-1}$ ($\beta=T^{-1}$ is inverse temperature). Translating the behavior of $\{b_n\}$ to the dynamics of K-complexity, we find that $\K(t)$ will first grow exponentially with the Lyapunov exponent saturating the chaos bound~\cite{Maldecena2016,Maldacena2016bound,Parker2019}, and after a scrambling time of order $O(\beta\log\beta)$, $\K(t)$ switch to linear growth in time, with the velocity the same as if it's in infinite temperature. This is summarized schematically in figure~\ref{fig:schematic}(a), (b).

In section~\ref{sec:remarks on finite N} we remark on the finite $N$ corrections and in section~\ref{sec:conclusion}, we summarize our result and discuss some possible future directions.

\paragraph{Related works.}
In reference~\cite{Bhattacharyya2023,Erdmenger2023}, the authors studied the `\textit{state}' Krylov complexity generated by random matrix Hamiltonian. Namely, they expand $|\psi(t)\rangle=e^{-iHt}|\psi_0\rangle$ on the Krylov basis which is constructed from the Lanczos orthogonalization of $\{H^n|\psi_0\rangle\}$. Despite having the same name, in our work we actually study `\textit{operator}' Krylov complexity generated by Heisenberg evolution of RMT, which is a different quantity. In reference~\cite{Kar2022}, the authors considered an `energy-refined' version of Krylov complexity and its relation with RMT universality class. Though bearing similar motivations to us, they studied different quantities from ours.

\section{Review of Krylov complexity}
\label{sec:review on K complexity}
In this section, we briefly review the definition of Krylov complexity and construction of Krylov basis~\cite{Parker2019}. Experts on this topic may skip this section and directly dive to the next one with new results.

Given an Hermtian initial operator $O\in \mathcal{B}(\mathcal{H})$ which is an bounded operator on Hilbert space $\mathcal{H}$ with dimension $N$, its Heisenberg evolution is given by $O(t)=e^{iHt}Oe^{-iHt}$, where $H$ is the Hamiltonian governing the time evolution of the quantum systems. Since we are interested in the dynamics of $O(t)$, it is convenient to work in the vector space $\BH$ of operator, with dimension $N^2$. This is achieved through operator-to-state mapping: $O\rightarrow|O)$.

In order to have a notion of normalized and orthogonal vectors, we need to equip this operator linear space with an inner product structure $(A|B)$, defined as
\begin{equation}
(A|B)\equiv
\begin{cases}
\tr[A^{\dagger}B],\ & \beta=0 \\
\tr[e^{-\beta H/2}A^{\dagger}e^{-\beta H/2}B],\ & \beta\neq0
\end{cases}
\end{equation}
for infinite temperature and finite temperature respectively. The motivation for inserting thermal density matrix into finite temperature inner product is because it resembles the thermal Wightman two-point function~\cite{Parker2019}.

The Heisenberg evolution is also mapped to a superoperator, called Liouvillian $\mL$, defined as the commutators of Hamiltonian: $\mL O\equiv[H,O]$. One can easily show that such an inner product would render Livioullian a Hermitian super-operator, namely $(A|\mL B)=(\mL A|B)$. By Baker-Hausdoff-Campell formula, the time evolution is thus expanded in the basis of $\{|\mL^nO)\}$:
\begin{equation}
|O(t))=e^{i\mL t}|O)=\sum\limits_{n=0}^{\infty}\frac{(it)^n}{n!}|\mL^n O)
\end{equation}

As we claimed in section~\ref{sec:introduction}, the idea of Krylov complexity is that can quantify the complexity `beyond real space': the complexity in Hilbert space. Namely, we expect the growth of $\K(t)$ to characterize the process in which $O(t)$ explores more and more linearly independent basis in $\BH$ that it has not occupied before. Certainly $\{|\mL^{n} O)\}$ is linearly independent in general, however, they are neither normalized nor orthogonal. Therefore, to make this idea explicit, we need to perform an orthogonalization procedure on $\{|\mL^nO)\}$ and generate an orthonormal basis $\{|O_n)\}$.

This procedure is the Lanczos algorithm. Starting from $|O)$, define a normalized vector $|O_0)=(O|O)^{-1/2}|O)$. The second step is to define $|O_1)=b_1^{-1}\mL|O_0)$ where $b_1=(\mL O_0|\mL O_0)^{1/2}$. Then inductively define:
\begin{equation}
\begin{aligned}
|A_n)&=\mL|O_{n-1})-b_{n-1}|O_{n-2})\\
b_n&=(A_n|A_n)^{1/2}\\
|O_n)&=b_n^{-1}|A_n)
\end{aligned}
\end{equation}

The output of this algorithm is a set of orthonormal basis $\{|O_n)\}$ satisfying $(O_n|O_m)=\delta_{nm}$ and a set of real and positive coefficient $\{b_n\}$ called Lanczos coefficient. One also notice that since $O$ is Hermitian, then $|O_n)$ is also a Hermitian operator which can be expressed as a degree-$n$-polynomial of $\mL$ acting on $|O)$. 

One also notices that the matrix of super-operator $\mL$ written in the basis $\{|O_n)\}$ is a tridiagonal real and Hermitian matrix with zero diagonal entries and its off-diagonal entries equating $\{b_n\}$, namely:
\begin{equation}
L_{mn}=(O_m|\mL|O_n)=
\begin{bmatrix}
0 & b_1 & 0 & 0 &  \cdots\\
b_1 & 0 & b_2 & 0 &  \cdots\\
0 & b_2 & 0 & b_3 &  \cdots \\
0 & 0 & b_3 & 0&  \ddots\\
\vdots & \vdots &\vdots &\ddots &\ddots
\end{bmatrix}
\label{eq:tridiagonal matrix}
\end{equation}

We also briefly remark that the dimension of Krylov space (spanning of $\{|O_n)\}$) denoted as $K$, is of the same order but slightly smaller than the dimension of $\BH$. This is because Hamiltonian commutes with any function of itself. In other words, $\{|H^n)\}$ spans the kernel space of $\mL$. Therefore even if initial operator $|O)$ may have non zero occupation on $\text{ker}(\mL)$, all other $\{\mL^{n}|O),n\geq1\}$ are perpendicular to $\text{ker}(\mL)$. Therefore at most we have $K\leq N^2-\text{dim}(\text{ker}(\mL))+1$. A careful examination shows that~\cite{Rabinovici2021} $K\leq N^2-N+1$. Nevertheless, in this paper, we will not probe the physics near the edge of Krylov space.

Now we are ready to study the dynamics of $O(t)$ by expanding $|O(t))$ in the basis $\{O_n\}$ with coefficient $\varphi_n(t)$:
\begin{equation}
|O(t))=\sum\limits_{n=0}^{K-1}\varphi_n(t)|O_n)
\end{equation}
Then $\varphi_n(t)$ satisfies the following Schrodinger-like equation:
\begin{equation}
i\partial_t\varphi_n=-b_{n+1}\varphi_{n+1}-b_n\varphi_{n-1}
\end{equation}
which describes a single particle initially at the left-end and then starts hopping on the one dimensional (semi-infinite, if we imagine $K$ is large) Krylov chain with non-uniform position-dependent hopping coefficients $\{b_n\}$.

The Krylov complexity is then defined as the average position of this wave packet:
\begin{equation}
\K(t)\equiv\sum\limits_{n=0}^{K-1} n |\varphi_n(t)|^2
\end{equation}
This definition is very intuitive compared with our daily life experience of defining complexity: To handle a complex task (say, build a house), we try to decompose it into a sequence of simple tasks (say, adding bricks), which we manage to deal with step by step. Then we use the number of steps to denote the overall complexity (say, the number of bricks). Many complexity measures in quantum systems follow this spirit. For instance, in quantum computing we often use circuit complexity to quantify a complex quantum state $|\psi\rangle$, which is defined as the number of elementary gates that are needed to assemble into a large unitary rotating simple state $|\psi_0\rangle$(often a product state with zero entanglement entropy) to $|\psi\rangle$. For our case of Krylov complexity, the simple operation is just the Heisenberg commutator $\mL=[H,\cdot]$ and K-complexity counts the average number of operations we needed to make a simple $O$ to be as complex as $O(t)$. This analogy is summarized in table~\ref{table:compare K-complexity and circuit}.

\begin{table}[t]
    \scriptsize
    \centering
    \begin{tabular}{c|c|c|c|c}\hline
\ & Simple initial object & Simple operation & Complex final object & Complexity measure \\
\hline
\makecell[c]{Circuit \\complexity} & \makecell[l]{$|\psi_0\rangle$ \\product state} & \makecell[l]{$\{U_i\}$\\ sets of universal gates} & \makecell[l]{$|\psi\rangle $\\ highly entangled} & number of operations\\
\hline
\makecell[c]{Krylov\\ complexity} & \makecell[l]{$O$\\ local operator} & \makecell[l]{$\mL=[H,\cdot]$ \\Heisenber commutator} & \makecell[l]{$O(t)$\\ highly scrambled}& \makecell[c]{average number\\ of operation}\\
\hline
    \end{tabular}
    \caption{Analogy between circuit complexity (as a state complexity measure shown above, but can also be readily generalized to measure operator complexity) and Krylov complexity (as an operator complexity measure). The universal set of gates $\{U_i\}$ are often one-qubit gates and two-qubit gates. One choice is all single qubit rotation together with CNOT gate.}
    \label{table:compare K-complexity and circuit}
\end{table}

We also remark on the difference between circuit complexity and Krylov complexity. For the former, the choice of `simple operation', namely a set of universal one-qubit/two-qubit gates, has ambiguity. But for the latter, the choice of commutator is very natural, since the dynamics is generated by it in the first place.

Going back to $\K(t)$, let's find out the relation between $\{b_n\}$ and $\K(t)$ through some intuitive examples. The first example is that $b_n\equiv b$ is constant. In this case, the hopping coefficient is uniform so we expect the center of the wave packet travel in a constant velocity proportional to $b$. Therefore $K(t)$ growth linear in time.

The next example of interest is to consider $b_n=\alpha n+\gamma$. Before that, we first want to take the continuous limit of the wave equation. Define $\phi_n(t)=i^{-n}\varphi_n(t)$ to make $\phi_n(t)$ satisfy a real wave equation $\partial_t\phi_n(t)=-b_{n+1}\phi_{n+1}+b_n\phi_{n-1}$ with initial condition $\phi_n(0)=\delta_{n0}$. We define a continuous position variable $x=\epsilon n$, where $\epsilon$ plays the role of lattice constant. Then $\epsilon\rightarrow0$ limit results in the following continuous equation~\cite{Jian2021}:
\begin{equation}
(\partial_t+2\alpha x\partial_x+\alpha )\phi(x,t)=0
\end{equation}
Given initial distribution $\phi_0(x)$, we obtain the unique solution:
\begin{equation}
\phi(x,t)=e^{-\alpha t}\phi_0(xe^{-2\alpha t})
\end{equation}
This shows that up to an overall damping factor $e^{-\alpha t}$ which keeps the normalization of wavefunction the same, the initial wave packet is stretched with factor $e^{2\alpha t}$. Therefore the average position $\langle x\rangle_{(t)}=e^{2\alpha t}\langle x\rangle_{(0)}$ will also grow exponentially, indicating a $\K(t)\sim e^{2\alpha t}$.

We also notice that $\K(t)$ is fully determined by two-point auto-correlation function $C(t)\equiv(O|O(t))$. This is because $\K(t)$ is determined by $\{b_n\}$, which is in turn determined by a set of moments $\{\mu_{2n}\equiv(O|\mL^{2n}|O)\}$, the derivatives of $C(t)$ at $t=0$. It is also worth remarking the relation between $\{\mu_{2n}\}$ and $\{b_n\}$ is highly non-linear~\cite{Parker2019}:
\begin{equation}
\mu_{2n}=(L^{2n})_{00}=\sum\limits_{\text{path}}b_{i_1}b_{i_2}\cdots b_{i_{2n}}
\label{eq:from b to mu}
\end{equation}
where $L$ is the tridiagonal matrix of $\mL$ written in Krylov basis, defined in equation~(\ref{eq:tridiagonal matrix}). Here, a Dyck path is defined by starting from $n=0$ and return $n=0$ with exactly $2n$ steps. The number of such paths with $2n$ steps is given by the Catalan number $C_{n}=\frac{(2n)!}{(n+1)!n!}$. The relation from $\{\mu_{2n}\}$ to $\{b_n\}$ is also complicated~\cite{Parker2019}:
\begin{equation}
b_1^2\cdots b_n^2=\operatorname{det}(\mu_{i+j})_{0\leq i,j\leq n}
\label{eq:from mu to b}
\end{equation}

Given $\{b_n\}$, in order to gain some intuition about the scaling behavior of $\{\mu_{2n}\}$ wrt $n$, which is important for our discussion in later sections, there is an explicit inequality bound for $\mu_{2n}$ if we assume $b_n$ is non-decreasing with $n$:
\begin{equation}
C_n(b_1^2b_2^2\cdots b_n^2)\geq\mu_{2n}>(b_1^2b_2^2\cdots b_n^2)
\label{eq:2.12}
\end{equation}
This is because $(b_1^2b_2^2\cdots b_n^2)$ is the maximal weight among all Dyck paths. For physically relevant cases, $b_n$ schematically has the common form $b_n\sim \alpha n^{\delta}$. $\delta=1,\frac{1}{2},0$ respectively correspond to~\cite{Parker2019}: chaotic local systems at stage one; integrable systems at stage one; free systems at stage one or chaotic local systems at stage two. Therefore, considering $n$ to be large but finite, we may apply Stirling formula to equation~(\ref{eq:2.12}):
\begin{equation}
2\delta n\log n+\left(2\log2+2\log\alpha-2\delta\right)n+O(\log n)\geq\log\mu_{2n}>2\delta n\log n+\left(2\log\alpha-2\delta\right)n+O(\log n)
\label{eq:inequality bound}
\end{equation}
If we further assume that $\log\mu_{2n}=A_1  n\log n+A_2 n+O(\log n)$ with the same scaling, then the leading-order coefficient $A_1$ completely fix $\delta$ and next-leading-order coefficient $A_2$ provide bound on $\alpha$.

Ever since the concept of Krylov complexity has been proposed~\cite{Parker2019}, it has attracted considerable attention from various communities since it exhibits many universal behaviors in chaotic local systems described in detail in section~\ref{sec:introduction}. People studied this quantity in many contexts ranging from condensed matter models focusing on scrambling behaviors~\cite{Barbón2019,Dymarsky2020,Tianrui2020,Chen2021,Noh2021,Fabian2022,Kim2022,Hörnedal2022,Bhattacharjee2022,Sinong2022,Zhongying2022,Fan2022,Rabinovici2021,Rabinovici20221,Rabinovici20222,Erdmenger2023,Kar2022}, open quantum systems~\cite{Liu2023,Bhattacharya2022}, field theory~\cite{Dymarsky2021_CFT,Caputa2021,caputa2021geometry,Adhikari2022,ADHIKARI2023116263}, and holographic gravity~\cite{Avdoshkin2020,Barbón2020,Magán2020,MUCK2022115948,Banerjee2022,Kar2022,Jian2021}. Now we are ready to add one more analytically tractable example, in RMT, to this exciting field with flourishing development.

\section{Large $N$ limit of K-complexity in RMT}
\label{sec:Large N}
\subsection{Infinite temperature}
\label{sec:infinite twmperature}
In this section, we analytically calculate Krylov complexity of RMT in large $N$ limit at infinite temperature. Our strategy is to first calculate $\{\mu_{2n}\}$ using RMT techniques. Ideally one would apply equation~(\ref{eq:from mu to b}) to get an expression of $\{b_n\}$, but it seems hopeless. Our strategy is to focus on the scaling behavior of $\mu_{2n}$ for large but finite $n$, then appeal to equation~(\ref{eq:from b to mu}) and inequality~(\ref{eq:inequality bound}) to estimate the scaling behavior of $\{b_n\}$ with respect to $n$.

For readers' convenience, we first recall some definitions about Krylov complexity:
\begin{equation}
\mu_{2n}=(O|\mathcal{L}^{2n}|O),\ \mathcal{L}=[H,\cdot]
\end{equation}
where the inner product in infinite temperature is defined as $(A|B)=\tr\left[A^{\dagger}B\right]$. We also notice that $\mu_{2n+1}=0$ since $\mL^{2n+1}O$ is anti-Hermitian therefore its trace with Hermitian matrix $O$ is zero. Using identity:
\begin{equation}
\mathcal{L}^{2n}O=\left([H,\right)^{2n}O]=\sum\limits_{k=0}^{2n}C_{2n}^k(-1)^kH^{k}OH^{2n-k}
\end{equation}
we derive the formal equation for $\mu_{2n}$:
\begin{equation}
\mu_{2n}=\sum\limits_{k=0}^{2n}C_{2n}^k(-1)^k\tr[ OH^{k}OH^{2n-k}]
\label{eq:3.3}
\end{equation}
For simplicity, we consider the simplest type of RMT, which is an $N\times N$ random matrix from GUE (Gaussian Unitary Ensemble). In GUE, the probabilistic distribution of $H$ is symmetric under any unitary transformation: $P(H)=P(UHU^{\dagger})$. Therefore, the moment $\mu_{2n}(O)=\mu_{2n}(UOU^{\dagger}),\ \forall U$, which means that it finally depends only on the spectrum of $O$. 

We can compare the case in~\cite{Bhattacharyya2023,Erdmenger2023}, where the author there studied the \textit{state complexity} rather than \textit{operator complexity} generated by GUE Hamiltonian. In their case, the initial state independence exist after average: $|\psi\rangle\rightarrow U|\psi\rangle$ means all initial states give the same Lanczos coefficient.

To proceed, we use the above mentioned fact that $P(H)=P(UHU^{\dagger})$, therefore we can do the following replacement:
\begin{equation}
\mathbb{E}\tr[OH^{k}OH^{2n-k}]=\E \tr[OUH^{k}U^{\dagger}OUH^{2n-k}U^{\dagger}],\ \forall U
\end{equation}
where notation $\E$ means taking average over GUE Hamiltonian.
Since the above equation is valid for arbitrary $U$, we can first keep $H$ fixed and average over $U$ under Haar measure, using Weingarten function~\cite{Roberts2017,Cotler2017chaos}:
\begin{equation}
\begin{aligned}
\int dU\ U_{i_1j_1}U_{i_2j_2}U^*_{i_3j_3}U^*_{i_4j_4}=&\frac{\delta_{i_1i_3}\delta_{i_2i_4}\delta_{j_1j_3}\delta_{j_2j_4}+\delta_{i_1i_4}\delta_{i_2i_3}\delta_{j_1j_4}\delta_{j_2j_3}}{N^2-1}\\
&-\frac{\delta_{i_1i_3}\delta_{i_2i_4}\delta_{j_1j_4}\delta_{j_2j_3}+\delta_{i_1i_4}\delta_{i_2i_3}\delta_{j_1j_3}\delta_{j_2j_4}}{N^3-N}
\end{aligned}
\end{equation}
Then we arrive at the following useful formula:
\begin{equation}
\begin{aligned}
\int dU\cdot \tr[OUAU^{\dagger}OUBU^{\dagger}]=\frac{1}{N^2-1}\bigg(&\tr[O^2]\tr[A]\tr[B]+\tr[O]^2\tr[AB]\\
&-N^{-1}\tr[O]^2\tr[A]\tr[B]-N^{-1}\tr[O^2]\tr[AB]\bigg)
\end{aligned}
\end{equation}
For simplicity we can consider a traceless normalized operator with $\tr[O]=0,\ \tr[O^2]=(O|O)=1$, then the moment $\mu_{2n}$ is given by:
\begin{equation}
\begin{aligned}
\E\ \mu_{2n}&=\E\ \frac{1}{N^2-1}\sum\limits_{k} C_{2n}^k(-1)^k\bigg(\tr[H^k] \tr[H^{2n-k}]-N^{-1}\tr[H^{2n}]\bigg)\\
&\approx \E\  N^{-2}\sum\limits_{k} C_{2n}^k(-1)^k\tr[H^k] \tr[H^{2n-k}]\\
&=\E N^{-2}\sum_{i,j}(\lambda_1-\lambda_j)^{2n}
\label{eq:mu}
\end{aligned}
\end{equation}
where $\lambda_{i}$ is the eigenvalue of $H$. For later convenience, in the second line, we perform approximation by only retaining the leading order terms in large $N$ limit. 

As a side remark, the second line of equation~(\ref{eq:mu}) can be derived using another approach. Instead of insisting that $O$ is a fixed operator, we can otherwise consider $O$ to be also a random Hermitian matrix under independent GUE distribution from $H$:
\begin{equation}
    \overline{O_{ij}O_{mn}}=N^{-2}\delta_{in}\delta_{jm}
\end{equation}
where the variance $N^{-2}$ is determined by normalization condition:
\begin{equation}
\overline{(O|O)}=\overline{\tr[O^2]}=\sum\limits_{i,j=1}^{N} \overline{O_{ij}O_{ji}}=1
\end{equation}
With the GUE distribution of $O$ we have identity $\overline{\tr[OAOB]}=N^{-2}\tr [A]\tr [B]$. Applying this to equation~(\ref{eq:3.3}), we have:
\begin{equation}
\overline{\mu_{2n}}=N^{-2}\sum\limits_{k} C_{2n}^k(-1)^k\tr[H^k] \tr[H^{2n-k}]
\end{equation}
which is similar to equation~(\ref{eq:mu}).

Nevertheless, we will proceed by adopting that $O$ is a fixed operator. Starting from equation~(\ref{eq:mu}), we see that $\E\mu_{2n}$ is identically the same for any traceless initial operator $O$, and only depend on the spectrum of $H$:
\begin{equation}
\begin{aligned}
\E\ \mu_{2n}&=\E\ N^{-2}\sum\limits_{k} C_{2n}^k(-1)^k\tr[H^k] \tr[H^{2n-k}]\\
&=\E\ N^{-2}\sum\limits_{k,i,j}C_{2n}^k(-1)^k \lambda_i^k\lambda_j^{2n-k}\\
&=\E\ N^{-2}\sum\limits_{i,j}(\lambda_i-\lambda_j)^{2n}\\
&=N^{-2}\int d\lambda d\lambda'\cdot (\lambda-\lambda')^{2n} \rho(\lambda,\lambda')
\end{aligned}
\end{equation}
where $\rho(\lambda,\lambda')$ is the averaged two-point spectral density of RMT. 

\begin{figure}[t]
    \centering
    \includegraphics[width=0.49\textwidth]{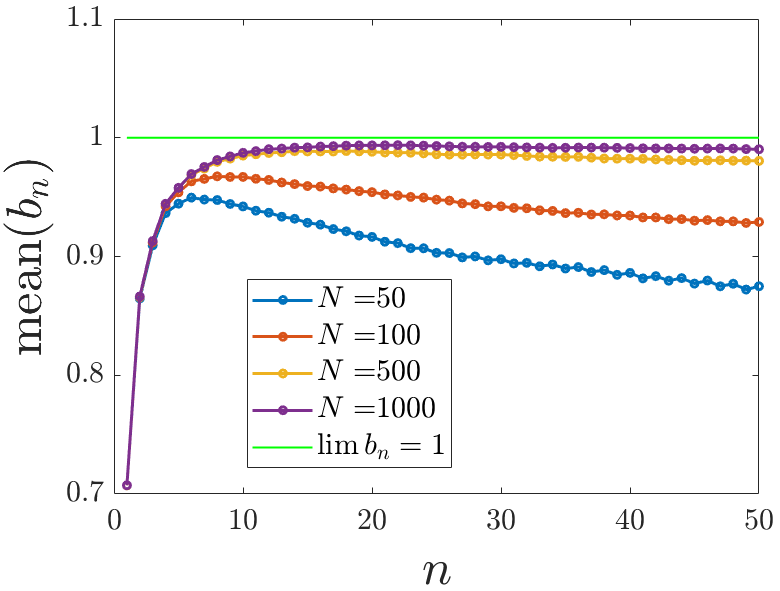}
    \includegraphics[width=0.49\textwidth]{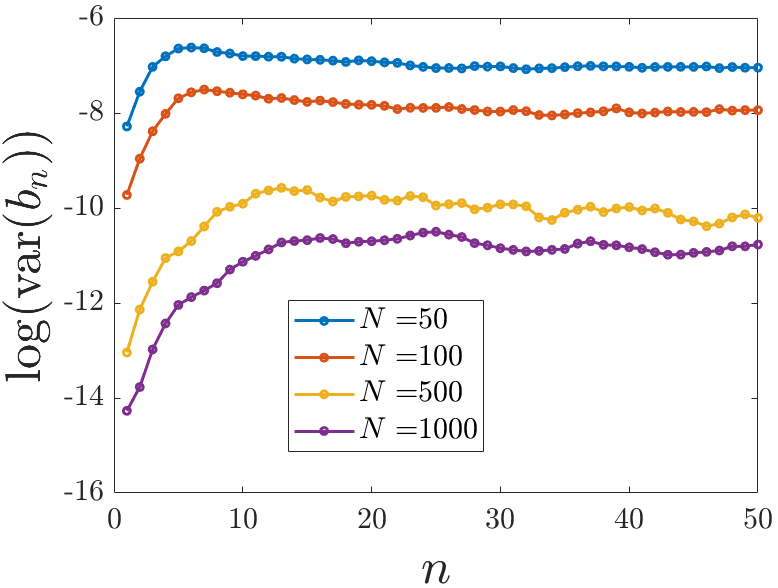}
    \caption{Numerical simulation of random matrix with GUE (Gaussian Unitary Ensemble) with 2000 samples. \textbf{Left:} The average value of $b_n$. \textbf{Right:} The log of variance. We observe that the variance of $b_n$ decreases roughly in power law wrt $N$. Therefore in large $N$ limits the variance is suppressed and reduces possible Anderson localization due to disorder.}
    \label{fig:simulation infinite temperature}
\end{figure}

Here we shall pause for a moment and make two relevant remarks.
\begin{itemize}
\item[1.] \textit{Order of taking random average}: The order of average we take here is the same as that when people study Krylov complexity in SYK model~\cite{Maldecena2016,Sachdev2015}. There, people first calculated random averaged two-point auto-correlation function $\E\ C(t)=\E (O|O(t))$. Then by taking derivatives we arrive at disorder averaged $\E \mu_{2n}$. The Krylov coefficient $b_n=f(\E\mu)$ is obtained subsequently using equation~(\ref{eq:from mu to b}), which we schematically denote as $f(\cdot)$. Of course, we can directly perform average over $\E b_n$, and since the fluctuation in RMT is controllably small in large $N$ limit, we expect the order of average to not affect the mean value,i.e., $\E f(\mu)=f(\E\mu)$, even if $f(\cdot)$ is a highly nonlinear map. Nevertheless, in the right panel of  figure~\ref{fig:simulation infinite temperature}, we evaluate the statistical variance of $\{b_n\}$ from numerical simulation of random matrix, and indeed verified that the variance is suppressed by some positive power of $N^{-1}$. Therefore, for notational symplicity, we will sometimes drop the average symbol $\E$.

\item[2.] \textit{Average over initial operator $O$}: In some recent papers\cite{Jafferis20231,Jafferis20232}, people find that if the ensemble of $O$ and $H$ are sampled independently, the averaged $n$-point correlator of $O$ cannot reproduce the known results of RMT for $n\geq4$. Here in our story, every data comes from the two-point auto-correlation function of $O$, therefore this possible subtlety will not show up in our story. Nevertheless, to avoid any possible ambiguity, we will proceed by adopting that $O$ is a fixed operator.
\end{itemize}

Going back to the main theme, we shall proceed by first introducing some notation and knowledge in random matrix theory. The spectral one point function $\rho(\lambda)$ and two point function $\rho(\lambda,\lambda')$ in GUE is defined by:
\begin{align}
&\rho(\lambda)=\E\sum_i\delta(\lambda-\lambda_i)=R_1(\lambda)\\
&\rho(\lambda,\lambda')=\E\sum_{i,j}\delta(\lambda-\lambda_i)\delta(\lambda'-\lambda_j)=\delta(\lambda-\lambda')R_1(\lambda)+R_2(\lambda,\lambda')
\end{align}
where $R_1,R_2$ are one-point and two-point spectral correlators, following the standard notation in RMT~\cite{mehta2004random}. For spectrum one point function $R_1(\lambda)$, it is given by the famous Wigner semicircle in large $N$ limit:
\begin{equation}
R_1(\lambda)=
\begin{cases}
2\pi^{-1}N\sqrt{1-\lambda^2} & \text{for}\ |\lambda|<1 \\
0 & \text{for}\ |\lambda|>1
\end{cases}
\end{equation}
where we scaled the probability distribution of GUE to be $P(H)\propto \exp(-2N\tr H^2)$ in order to set the radius of Wigner semicircle to be unity.

For spectrum two-point function $R_2(\lambda,\lambda')$, it has disconnected part $R_1(\lambda)R_2(\lambda')$ and connected part $T_2(\lambda,\lambda')$ related by:
\begin{equation}
R_2(\lambda,\lambda')=R_1(\lambda)R_1(\lambda')-T_2(\lambda,\lambda')
\end{equation}

The generic result for $T_2(\lambda,\lambda')$ is complicated, and its simplification appears only when we zoom into the center of the Wigner circle with a few windows with width determined by mean level spacing $O(N^{-1})$, i.e., we take the limit where $N\rightarrow+\infty,N\lambda\rightarrow \text{Const},N\lambda'\rightarrow \text{Const}$:
\begin{equation}
T_2(\lambda,\lambda')\approx(2\pi^{-1}N)^2\left[\frac{\sin 2N(\lambda-\lambda')}{2N(\lambda-\lambda')}\right]^2,\ \text{when}\  N\rightarrow+\infty,N\lambda,N\lambda'\rightarrow\text{Const}
\end{equation}
Plugging everything into the expression we have:
\begin{equation}
\mu_{2n}\approx4\pi^{-2}\int d\lambda d\lambda'\cdot(\lambda-\lambda')^{2n}\left\{\sqrt{1-\lambda^2}\sqrt{1-\lambda'^2}-\left[\frac{\sin 2N(\lambda-\lambda')}{2N(\lambda-\lambda')}\right]^2\right\}    
\end{equation}

We first omit the connected part in large $N$ limit. So the leading order in large $N$ would be:
\begin{equation}
\begin{aligned}
\mu_{2n}&\approx4\pi^{-2}\int_{-1}^{1} d\lambda d\lambda'\sqrt{1-\lambda^2}\sqrt{1-\lambda'^2}(\lambda-\lambda')^{2n}\\
&=4\pi^{-2}\int_{0}^{\pi}d\theta d\theta'\sin^2\theta\sin^2\theta'(\cos\theta-\cos\theta')^{2n}\\
&=4\pi^{-2}\sum\limits_{k=0}^{2n} C_{2n}^k(-1)^k\left[\int_{0}^{\pi}d\theta\sin^2\theta\cos^k\theta\right]\left[\int_{0}^{\pi}d\theta\sin^2\theta\cos^{2n-k}\theta\right]\\
&=4\pi^{-2}\sum\limits_{k=0}^{2n} C_{2n}^k(-1)^k\left[F_{k}-F_{k+2} \right]\left[F_{2n-k}-F_{2n-k+2} \right]
\end{aligned}
\end{equation}
where the integral $F_{k}$ is elementary:
\begin{equation}
F_k=\int_{0}^{\pi}d\theta\cos^{k}\theta=
\begin{cases}
\frac{\Gamma\left(\frac{1}{2}\right)\Gamma\left(\frac{k}{2}+\frac{1}{2}\right)}{\Gamma\left(\frac{k}{2}+1\right)} & \text{for}\ k\in \text{even}\\
0 & \text{for}\ k\in \text{odd}
\end{cases}
\end{equation}
Plugging the expression of $F_k$, then we arrived at the analytical form of $\mu_{2n}$ in leading order of large $N$:
\begin{equation}
\mu_{2n}=2\pi^{-1}\sum\limits_{m=0}^{n}\frac{(2n)!}{(2m)!(2n-2m)!}\frac{\Gamma\left(m+\frac{1}{2}\right)\Gamma\left(n-m+\frac{1}{2}\right)}{\Gamma\left(m+2\right)\Gamma\left(n-m+2\right)}
\end{equation}
It seems hopeless to obtain a closed expression for this summation, therefore we try to seek the asymptotic behavior of $\mu_{2n}$ when $n\gg1$. To do this, we appeal to the Stirling formula $\log\Gamma(z)\approx z\log z-z+O(\log z)$:
\begin{equation}
\begin{aligned}
\mu_{2n}&\approx2\pi^{-1}\sum\limits_{m=0}^{n}\exp\left\{ -n\left[2x\log x+2(1-x)\log(1-x)\right]-3\log n-\frac{3}{2}\log x-\frac{3}{2}\log (1-x) \right\}\\
&\approx2\pi^{-1}n^{-2}\int_{-1}^1dx\exp\left(-nf(x)+O(\log n)\right),\ \text{where}\ x\equiv\frac{m}{n},\ f(x)\equiv 2x\log x+2(1-x)\log(1-x)\\
&\approx2\pi^{-1}n^{-2}\int_{-\infty}^{+\infty}dx\exp\left[-nf\left(1/2\right)-\frac{1}{2}nf''\left(1/2\right)(x-1/2)^2\right]\\
&\approx\exp\left[(2\log2)n+O(\log n)\right]
\end{aligned}
\end{equation}
In the first line, we use Stirling formula; in the second we approximate summation by integral; in the third line we perform integral using saddle point approximation, namely expanding around minimal of $f(x)$ and perform Gaussian integral, and in the last line we keep contribution only up to order $O(n)$ on the exponential. This is because, in the first line of approximation, the precision is already only up to order $n$, since if we want to keep to $O(\log n)$, we should use a finer Stirling formula $\log\Gamma(z)\approx z\log z-z-\frac{1}{2}\log z+O(1)$ 

To summarize, we have an exponential scaling behavior of $\mu_{2n}$:
\begin{equation}
\log\mu_{2n}\approx n\cdot2\log2+O(\log n)
\label{eq:3.23}
\end{equation}
Next, we notice that this exponential behavior of $\mu_{2n}$ would translate to the Constant behaviour of $b_n\equiv b$ for large $n$. From equation~(\ref{eq:from b to mu}), we see that if $b_n$ equals constant, we only need to count the number of paths:
\begin{equation}
\mu_{2n}=C_n b^{2n}=\frac{(2n)!}{(n+1)!(n)!}b^{2n}
\end{equation}
where $C_n$ is the Catalan number which counts the number of paths~\cite{Parker2019}. In order to compare the scaling behaviour of $\mu_{2n}$ for large but finite $n$, we again appeals to Stirling formula:
\begin{equation}
\log\mu_{2n}\approx{n\cdot2\log2+2n\log b+O(\log n)}
\end{equation}
\begin{equation}
    \Longrightarrow b=1
\label{eq:3.26}
\end{equation}
This suggests that $b_n\approx1$ will approach a constant value when $n$ is large!

We confirm this analytical prediction by direct numerical simulation of GUE random matrices, as presented in the left panel of figure~\ref{fig:simulation rescaling of H}. Numerically we see that $b_n$ is already close enough to 1 even when $n=10$. Since there is no scale at all for RMT at large $N$ and infinite temperature, we may well represent $b_n=1$ for $n$ larger than some order one value. This is the reason why we draw the behavior of $b_n$ as a horizontal straight line in the schematic plot in figure~\ref{fig:schematic}(a).

As we have already pointed out in section~\ref{sec:review on K complexity}, $b_n$ equating constant indicates that in large $N$ limit, $\K(t)$ in RMT grows linearly in time, in sharp contrast
to the chaotic quantum system in thermodynamic limit, whose K-complexity grow exponentially in time, as mentioned in section~\ref{sec:introduction}. We remark that this
does not mean K-complexity in RMT is smaller than that of chaotic local systems. This
is because, in RMT theory, we usually scale the Hamiltonian such that its eigenenergy is
of order $O(1)$. But in chaotic local systems, the energy is an extensive value which is of
order $O(S)$, where $S$ is the number of degree of freedom. Since the energy scale is conjugate to the time scale, we should properly rescale our
RMT result in order to compare with chaotic local systems.

We also remark that only the fact that it approaches constant is important, not the specific value of that constant. One can show directly from the Lanczos algorithm with infinite temperature inner product, that $b_n$ is proportional to the overall scaling of Hamiltonian. So, here deriving $b_n$ to approach 1 is simply a coincidence that we scale the radius semicircle to unity.

As a side remark, to make sure the linear growth of complexity on constant-$b$-plateau, we need to ensure that the fluctuation of $b_n$ is small enough in order to free from Anderson localization on Krylov chain~\cite{Rabinovici20221,Rabinovici20222}. Luckily, this is also satisfied in large $N$ limit, where in the right panel of figure~\ref{fig:simulation infinite temperature} we show that the statistical variance of $\{b_n\}$ is suppressed by some positive power of $N^{-1}$.

\subsection{Numerically comparing with generic chaotic local systems.}
\label{sec:numerical compare chaotic}

\begin{figure}[t]
    \centering
    \includegraphics[width=0.49\textwidth]{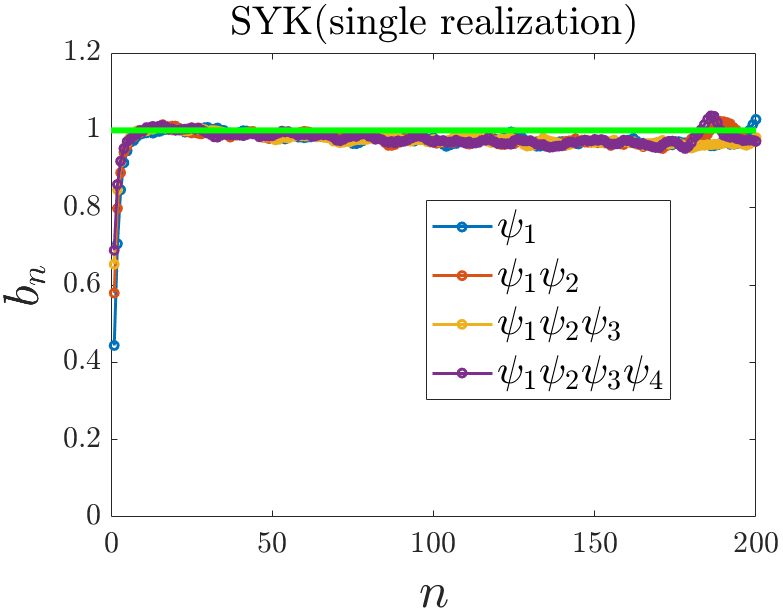}
    \includegraphics[width=0.49\textwidth]{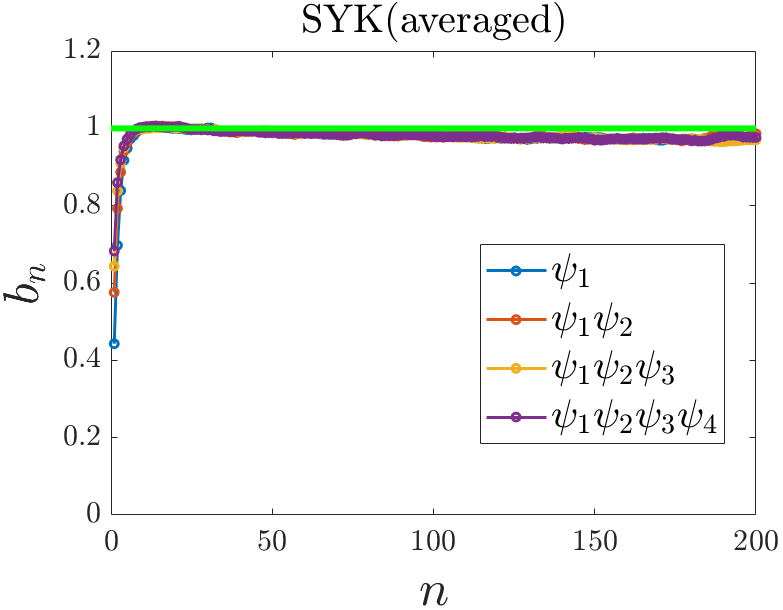}
    \includegraphics[width=0.49\textwidth]{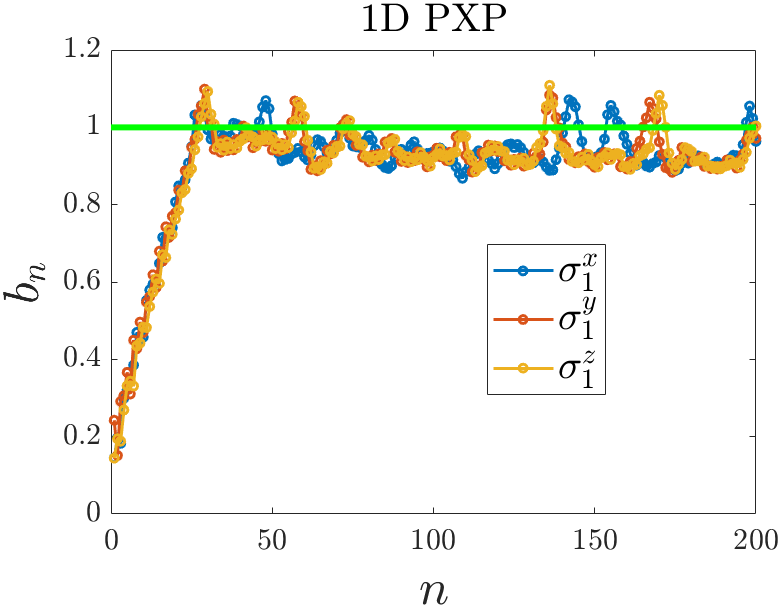}
    \includegraphics[width=0.49\textwidth]{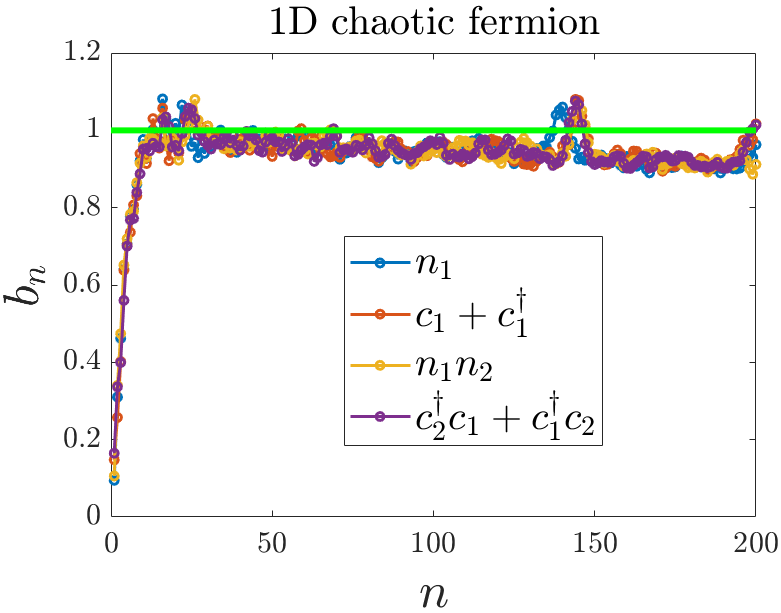}
    \includegraphics[width=0.49\textwidth]{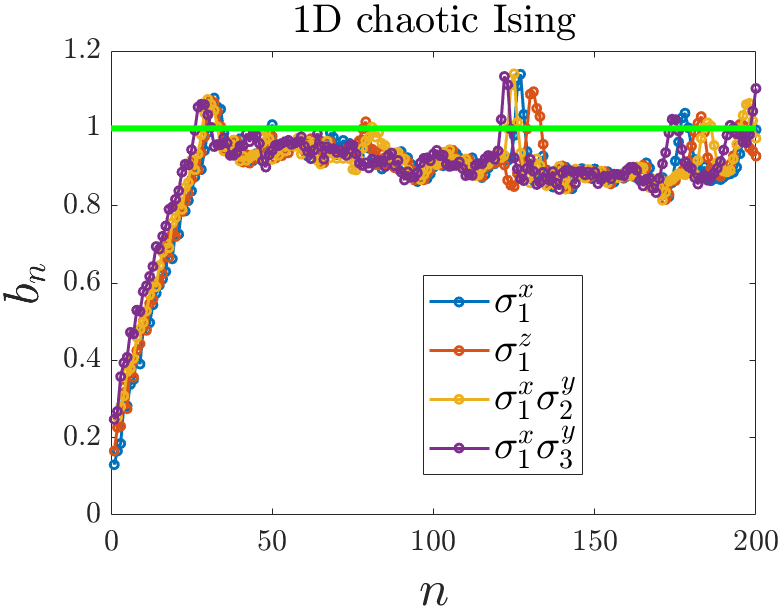}
    \includegraphics[width=0.49\textwidth]{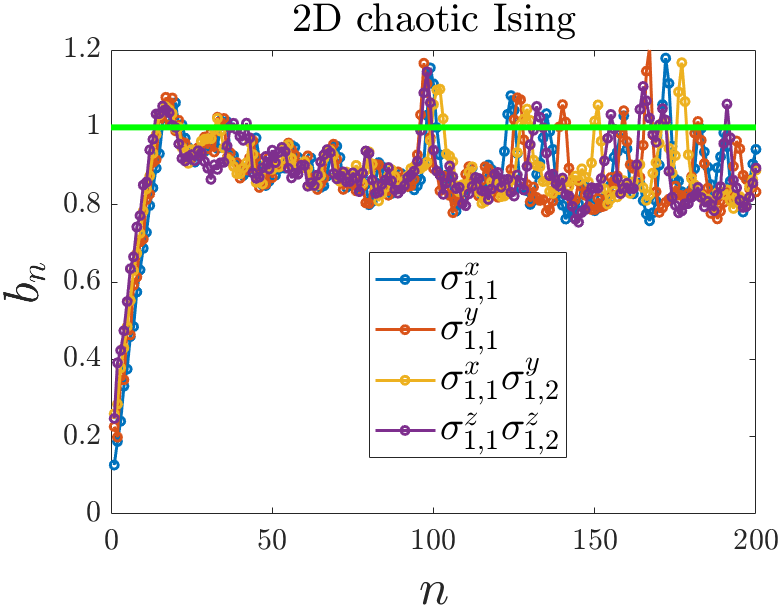}
\caption{Numerical simulation of Lanczos coefficient of $\{b_n\}$ in various chaotic local systems. In the same panel, different color lines correspond to different initial operators $O$ (we make them normalized and traceless). $b_n$ is shown in the unit of $(E_{\max}-E_{\min})/2$, where $E_{\max}$ and $E_{\min}$ are the maximal and minimal eigenvalues of Hamiltonian of each panel. For SYK model in equation~(\ref{eq:H_syk}), we take $J=1$ and $N_{\text{majorana}}=20$; for PXP model in equation~(\ref{eq:H_PXP}), we take $\Omega=1, U=-0.5$ on 16-sites; for chaotic fermion model in equation~(\ref{eq:H_fermion}), we take $t_1=1,t_2=0.64,V_1=0.88,V_2=0.76$ on 10-sites; for Ising model in equation~(\ref{eq:H_ising}), we take $g=1.87,h=1.92$ with 10-sites for 1D and $3\times3$ sites for 2D.}
    \label{fig:chaotic simulation}
\end{figure}

As explained in section~\ref{sec:introduction}, one motivation for studying Krylov complexity in RMT is that for a generic chaotic Hamiltonian with locality, the complexity growth after scrambling time $t_*$ is governed by random matrix behavior. This is because, at $t_*$, $O(t)$ has just finished its ‘scrambling in real space’, and begun to start its ‘scrambling in Hilbert space’. Therefore in stage two, $O(t)$ completely abandoned any spatial locality structure, or equivalently, the local Hamiltonian $H$ no longer looks local from
the perspective of $O(t)$. Therefore, in this stage, $H$ would act like a chaotic Hamiltonian without spatial locality, resembling a typical random matrix.

In this section, we numerically study the Lanczos coefficient $\{b_n\}$ of various initial operator $O$ in several chaotic local systems in various dimensions, including Ising model in one or two dimensions with both transverse field and longitudinal field, SYK model~\cite{Maldecena2016,Sachdev2015}, one-dimensional spinless fermion model with nearest-neighbor and next-nearest-neighbor hopping and density-density interaction, and PXP model as an effective model describing blockated Rydberg atom array~\cite{Bernien2017}. Their Hamiltonian is explicitly given by:
\begin{equation}
H_{\text{Ising}}=-\sum\limits_{\langle ij\rangle}\sigma_i^x\sigma_j^x+\sum\limits_i\left(g\sigma^z_i+h\sigma_i^x\right)
\label{eq:H_ising}
\end{equation}
\begin{equation}
H_{\text{SYK}}=\sum\limits_{i<j<k<l}J_{ijkl}\chi_i\chi_j\chi_k\chi_l,\ \overline{J^2_{ijkl}}=\frac{(4-1)!}{N_m^{4-1}}J^{2}
\label{eq:H_syk}
\end{equation}
\begin{equation}
H_{\text{fermion}}=-\sum\limits_i\left(t_1c_{i+1}^{\dagger}c_i+t_2c_{i+2}^{\dagger}c_i+\text{h.c.}\right)+\sum\limits_i\left(V_1n_{i+1}n_{i}+V_2n_{i+2}n_{i}\right)
\label{eq:H_fermion}
\end{equation}
\begin{equation}
H_{\text{PXP}}=P\left[\sum\limits_i\left(\Omega\sigma_i^x+U\sigma_i^z\right)\right]P
\label{eq:H_PXP}
\end{equation}
In SYK model, we use $N_m$ to denote the number of Majoranas. In PXP model, $P$ is a projection operator on to the subspace that any nearest neighbor sites cannot be both spin up.

To compare with RMT result, we should properly rescale $H_{\text{RMT}}$ to $\eta S\cdot H_{\text{RMT}}$ such that $\eta S\sim O(S)$. Therefore the plateau value of $b_n$ would rescale to $\eta S$, which is the same order as $b_n$ from $H_{\text{chaotic}}$. This rescaling also shifts the radius of Wigner semicircle from unity to $\eta S$, the same order as the extensive energy from $H_{\text{chaotic}}$. However, the order one coefficient $\eta$ is model-dependent and exhibits ambiguity. In order to obtain a model-independent rescaling ansatz, we define the rescaling factor to be such that the diameter of Wigner semicircle equals the width of the spectrum of $H_{\text{chaotic}}$. Therefore, the plateau value of RMT is scaled to be $b=(E_{\max}-E_{\min})/2$. 

Numerical results in figure~\ref{fig:chaotic simulation} shows that up to some erratic fluctuations, $b=(E_{\max}-E_{\min})/2$ is indeed slightly larger than $\{b_n\}$ from $H_{\text{chaotic}}$, with SYK model almost saturate RMT result. Therefore, we may carefully conjecture that in any chaotic local quantum systems, the infinite temperature Lanczos coefficient $\{b_n\}$ is bounded by $b_n\leq \eta_0(E_{\max}-E_{\min})/2$, with $\eta_0\sim O(1)$ a model-independent constant.

\subsection{Finite temperature}
\label{sec:finite temeprature}
For finite temperature, recall the inner product is defined by Wightman function:
\begin{equation}
(A|B)=\tr\left[e^{-\beta H/2}A^{\dagger}e^{-\beta H/2}B\right]
\end{equation}
Similar calculation gives the expression of $\mu_{2n}$:
\begin{equation}
\mu_{2n}=N^{-2}\sum\limits_{i,j}(\lambda_i-\lambda_j)^{2n}e^{-\beta(\lambda_i+\lambda_j)/2}
\label{eq:mu, finite temperature}
\end{equation}

\begin{figure}[t]
    \centering
    \includegraphics[width=0.53\textwidth]{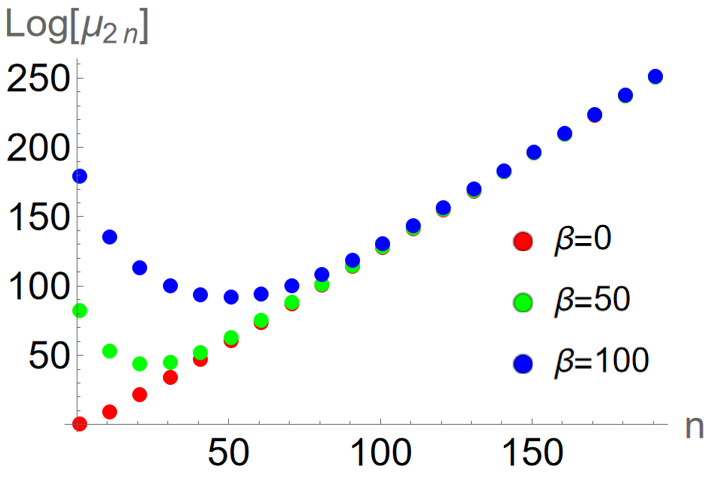}
    \includegraphics[width=0.45\textwidth]{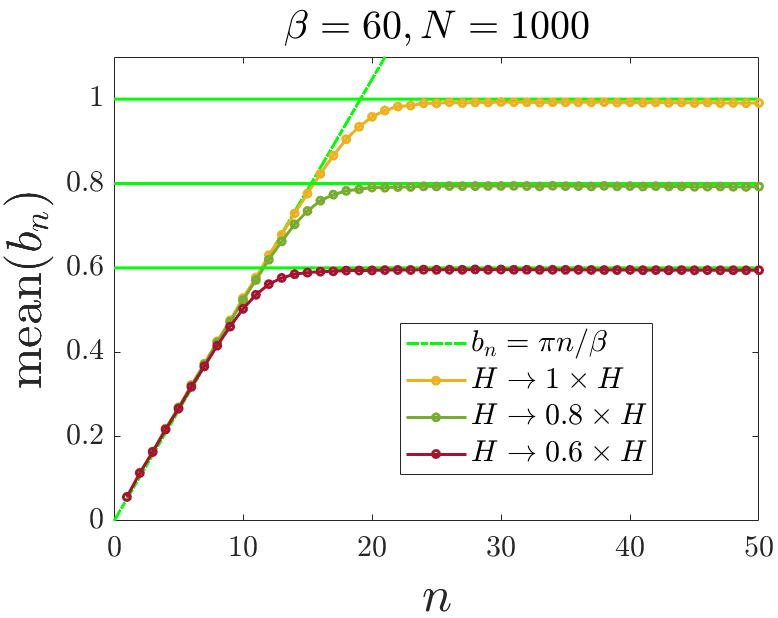}
    \caption{\textbf{Left:} $\mu_{2n}$-dependence on $n$, calculated at infinite $N$ limit, from numerical integrating first line of equation~(\ref{eq:nu_2n^(0)}). \textbf{Right:} Numerical simulation of $b_n$ in RMT ($N=1000$ is sufficient to represent infinite $N$ limit for our purposes), showing that the overall rescaling of Hamiltonian only proportionally changes the plateau value of $b_n$, while keeping the slope of initial linear-in-$n$ region intact.}
    \label{fig:simulation rescaling of H}
\end{figure}
Using the same RMT technique in large $N$ limit:
\begin{equation}
\begin{aligned}
\mu_{2n}&=4\pi^{-2}\int_{-1}^{1}d\lambda d\lambda'\sqrt{1-\lambda^2}\sqrt{1-\lambda'^2}(\lambda-\lambda')^{2n}e^{-\beta(\lambda+\lambda')/2}\\
&=4\pi^{-2}\sum\limits_{k=0}^{2n} C_{2n}^k(-1)^k\left[F_{k}-F_{k+2} \right]\left[F_{2n-k}-F_{2n-k+2} \right]
\end{aligned}
\label{eq:nu_2n^(0)}
\end{equation}
where the integral $F_{k}$ is given by the derivative of the modified Bessel function of the first kind:
\begin{equation}
F_{k}=\int_0^{\pi}d\theta \cos^k\theta e^{-\beta\cos\theta/2}=\pi(-1)^k I_0^{(k)}(\beta/2)
\end{equation}
with $I_{0}(z)=\pi^{-1}\int d\theta e^{-z\cos\theta}$ is the zeroth order of modified Bessel function of first kind, and $I_0^{(k)}(z)$ denotes its $k$-th derivative. Using the property that $I_0(z)=I_0(-z)$ and the identity $\partial^{n}(fg)=\sum C_n^k f^{(k)}g^{(n-k)}$, we can perform summation over $k$ and arrive at a simplified result:
\begin{equation}
\mu_{2n}=4\partial_{x}^{2n}\left[\left(I_0(x)-I_0^{(2)}(x)\right)^{2}\right]\bigg|_{x=\beta/2}
\end{equation}
Of course one can use the expansion $I_0(z)=\sum\frac{1}{(k!)^2}(z/2)^{2k}$ to study the behaviour of $\nu_{2n}$ in general temperature, but to get a feeling, we can study low temperature region when $\beta$ is large. We use the asymptotic behaviour of Bessel function:
\begin{equation}
I_0(x)\sim \frac{e^{x}}{\sqrt{2\pi x}},\ x\gg1
\end{equation}
So we have $(I_0(x)-I_0^{(2)}(x))^2\propto e^{2x}x^{-3}\sim e^{2x}$, where we ignored the $\log \beta$ correction on the exponential. Therefore:
\begin{equation}
\mu_{2n}\sim\partial_{x}^{2n}\left[e^{2x}\right]\bigg|_{x=\beta/2}=2^{2n}e^{\beta}\Longrightarrow\mu_{2n}\sim\exp(n\cdot2\log2)
\end{equation}
Interestingly, under this simple approximation, one may observe that finite temperature seemingly doesn't affect the scaling behaviour of $n$, compared to infinite temperature. This also means $b_n=1=\text{Const}$, following the same argument starting from equation~(\ref{eq:3.23}) to~(\ref{eq:3.26}).

\begin{figure}[t]
    \centering
    \includegraphics[width=0.49\textwidth]{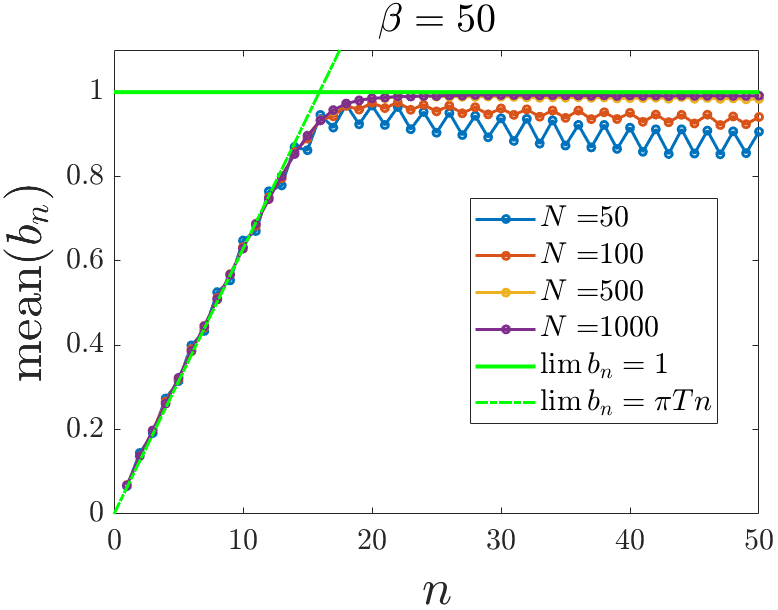}
    \includegraphics[width=0.49\textwidth]{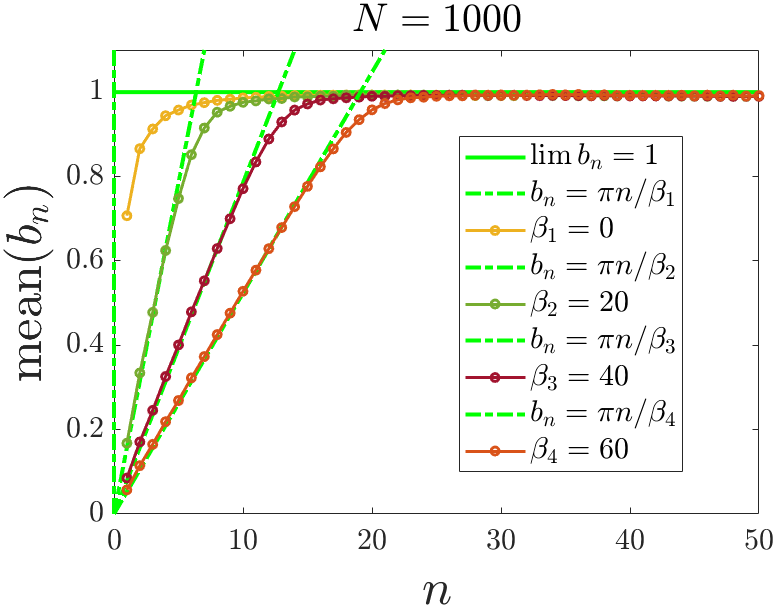}
    \caption{Numerical simulation of random matrix from GUE using 1000 samples. \textbf{Left:} We fix temperature, and see how $b_n$ behaves wrt $N$. In large $N$ limits, we see that $b_n$ first linearly increases with $n$ and then saturates to the platform of $b_n=1$. \textbf{Right:} $N=1000$ is sufficiently large to represent infinite $N$ result for our purposes. We find that in low temperature limit, the slope of initial linear-in-$n$ growth is precisely bounded by and saturates the chaos bound.}
    \label{fig:finite temperature simulation}
\end{figure}

Although the above calculation is in low temperature limit, we check numerically the exponential scaling behaviour of $\mu_{2n}$ actually appears for finite $\beta$ and large $n$, namely we first take $n$ goes to infinity while keeping $\beta$ finite, see figure~\ref{fig:finite temperature simulation}.

We also numerically calculate $b_n$ by direct simulation of RMT. The result is shown in figure~\ref{fig:finite temperature simulation}. We also confirm that the $b_n=1$ region appears at large $n$ and finite $\beta$. For finite $n$ but large $\beta$, there is a region where $b_n$ linearly increases with $n$.

In order to explain this linear growing region of $b_n$, we directly consider the integral of $\mu_{2n}$:
\begin{equation}
\mu_{2n}=4\pi^{-2}\int_{0}^{\pi}d\theta d\theta'\cdot\sin^2\theta\sin^2\theta'(\cos\theta-\cos\theta')^{2n}e^{-\beta(\cos\theta+\cos\theta')/2}
\end{equation}
We want to evaluate this integral by taking low temperature  ($\beta\rightarrow\infty$). In this way we change dummy variable $\varphi=\pi-\theta,\varphi'=\pi-\theta'$ and expand the integrand for small $\varphi,\varphi'$:
\begin{equation}
\mu_{2n}\approx2\pi^{-2}e^{\beta}2^{-2n}\int_{-\pi}^{\pi}d\varphi d\varphi'(\varphi^2-\varphi'^2)^{2n}e^{-\beta(\varphi^2+\varphi'^2)/4}
\end{equation}
where we also extend the integral range from $[0,\pi]$ to $[-\pi,\pi]$ using inversion symmetry of integrand. For low temperature, this Gaussian distribution is located at zero with width $O(\beta^{-1/2})$, parametrically small compared to integral range $[-\pi,\pi]$. Therefore we can safely extend the integral range to infinity. By changing variable to $\overline{\varphi}\equiv\varphi+\varphi',\delta\varphi=\varphi-\varphi'$, this Gaussian integral is evaluated as:
\begin{equation}
\begin{aligned}
\mu_{2n}&\approx\pi^{-1} e^{\beta}2^{-2n+2}\beta^{-1}\left(\left\langle\overline{\varphi}^{2n+4}\right\rangle_{\sigma}\left\langle\delta\varphi^{2n}\right\rangle_{\sigma}+\left\langle\overline{\varphi}^{2n}\right\rangle_{\sigma}\left\langle\delta\varphi^{2n+4}\right\rangle_{\sigma}-2\left\langle\overline{\varphi}^{2n+2}\right\rangle_{\sigma}\left\langle\delta\varphi^{2n+2}\right\rangle_{\sigma}\right)\\
&=\pi^{-1}\beta^{-2n-3}2^{3}e^{\beta}\left(\frac{(2n+2)!}{(n+1)!}\right)^2\frac{1}{2n+1}
\end{aligned}
\end{equation}
where $\langle\cdot\rangle_{\sigma}$ is the Gaussian average with variance $\sigma=4\beta^{-1}$. Going from the first line to the second line, we use $\langle z^{2n}\rangle_{\sigma}=\sigma^{n}\frac{(2n)!}{2^nn!}$.

Using Stirling formula for large but finite $n$, we find that the scaling behaviour of $\log\mu_{2n}$ with respect to $n$:
\begin{equation}
\log\mu_{2n}\approx 2n\log n+(4\log2-2+2\log\beta^{-1})n+O(\log n)
\end{equation}

Assuming ansatz $b_n=\alpha n^{\delta}$ and using inequality~(\ref{eq:inequality bound}), we can fix $\delta=1$ and provide a bound for $\alpha$:
\begin{equation}
4\beta^{-1}>\alpha>2\beta^{-1}
\end{equation}
This is indeed consistent with numerical simulation in the right panel of figure~\ref{fig:finite temperature simulation}, where we found $b_n=\pi\beta^{-1}n$, with $\alpha=\pi\beta^{-1}$, which is indeed within the bound.

We notice that from the continuous limit of the operator wave function on Krylov chain, as analyzed in section~\ref{sec:review on K complexity}, we immediately conclude that the Krylov complexity $\K(t)$ at low temperature will exhibit the following two-stage growth:
\begin{itemize}
\item[1.] \textit{Exponential growth before scrambling time.} $\K(t)\sim e^{\frac{2\pi}{\beta}t}$. We also notice that $\K(t)$ is interpreted as the average position of the wave function on the semi-infinite Krylov chain, therefore this exponential growth will proceed until the average position reaches the plateau, where the linear growth of $b_n$ ceases, namely when $\K(t_*)\sim O(\beta)$. This shows the scrambling time scale $t_*$ is of order $O(\beta\log\beta)$.
\item[2.]\textit{Linear growth after scrambling time.} $\K(t)\sim t$. When $t\gtrsim t_*$, the average position of the wave packet will travel to the region where $b_n$ is constant. Therefore it becomes a usual wave packet traveling ahead with constant velocity.
\end{itemize}

We also notice that at low temperature and before scrambling time, the exponential growth of Krylov complexity saturates the chaos bound of Lyaponov exponent $\lambda_L=\frac{2\pi}{\beta}$~\cite{Maldacena2016bound}. 

On the one hand, this is natural since the original literature on Krylov complexity~\cite{Parker2019} has proved that K-complexity can bound many other complexity measures like OTOC and operator size at infinite temperature, and for finite temperature the authors there conjectured the same bound to be true and verified on SYK model. 

On the other hand, the initial linear grow of $b_n$ in RMT should come as a surprise, since the linear growing behavior of $b_n$, though generic in chaotic many body systems, actually originates from the locality of Hamiltonian~\cite{Parker2019}. But clearly a random matrix Hamiltonian lacks locality. One interpretation is that the spectrum of RMT at low energy, which scales as $\rho(E)\propto \sqrt{E}$ agrees with the spectrum of a quantum black hole in JT gravity (see reference~\cite{Mertens2023} for a recent review) or SYK model~\cite{Maldecena2016} in low energy, which scales as $\rho_{JT}(E)\propto\sinh(2\pi\sqrt{2CE})\sim\sqrt{E}$~\cite{Mertens2023}. Matching of low energy density of state suggests that RMT has similar behavior with black hole in the view of fast scrambler~\cite{YasuhiroSekino2008}.

To justify that this low temperature chaos bound is universal, one necessary condition is that the coefficient of exponential growth in time, equivalently the slope of $b_n$ wrt to $n$, should not change when we vary the overall scaling of Hamiltonian while keeping temperature to be fixed. We notice that this is not possible in infinite temperature case. We recall the fact that $\{b_n\}$ is simply the matrix elements of Liouville superoperator $\mathcal{L}=[H,\cdot]$ writing in an orthonormal Krylov basis $\{|O_n)\}$. Therefore, naively we should expect $\{b_n\}$ to be proportional to the overall scaling of $H$. This is indeed true in infinite temperature, since the inner product $(A|B)=\tr[A^{\dagger}B]$ used to normalize the basis $\{|O_n)\}$ do not depend on $H$, therefore one can show by induction that $\{|O_n)\}$ do not depend on the overall scaling of $H$. However, this is not the case for finite temperature, where we choose Wightman function to define the inner product $(A|B)_{\beta}=\tr[e^{-\beta H/2}A^{\dagger}e^{-\beta H/2}B]$, which explicitly depend on $H$. As a result the basis $\{|O_n)\}$ depends on the overall scaling of $H$ in a complicated way and $\{b_n\}$ no longer proportional to it.

From right panel of figure~\ref{fig:simulation rescaling of H} we find that for finite temperature, the constant plateau part of $\{b_n\}$ is indeed affected by overall scaling, resembling infinite temperature behavior, while the initial slope remains intact from overall scaling! This perfectly justified the fact that the saturation of RMT to chaos bound is not a coincidence.

One interesting thing is that for RMT in $N\rightarrow \infty$, we obtain a finite scrambling time $t_*^{\text{RMT}}\sim O(\beta\log\beta)$, which is in contrast with chaotic local quantum systems $t_*^{\text{chaotic}}\sim O(\beta\log S)$, diverging in thermodynamic limit $S\rightarrow\infty$. Similar to the situation discussed in section~\ref{sec:numerical compare chaotic}, the reason is merely the overall scaling of Hamiltonian. From the right panel of figure~\ref{fig:finite temperature simulation}, we can of course prolong the scrambling time by scaling $H_{\text{RMT}}\rightarrow  O(S)\cdot H_{\text{RMT}}$, in order to match the scrambling time. Since $S\sim\log N$, this rescaling is equivalent to changing the variance of RMT probability distribution: $\log P(H)\sim -2N\tr H^2\longrightarrow \log P(H)\sim-\frac{2N}{\log^2 N}\tr H^2$, which is strange.

In RMT with scale controlled by probability distribution $P(H)\propto \exp(-2N\tr H^2)$, the energy spectrum is of order $O(1)$, which remains finite when $N\rightarrow\infty$. For chaotic local quantum systems, it energy spectrum diverges since energy is extensive.

\section{Remarks on finite $N$ corrections}
\label{sec:remarks on finite N}
A general remark would be that the level repulsion in connected part would have more impact for $\mu_{2n}$ with larger $n$. This is because level repulsion would make $\lambda_i-\lambda_j$ larger. We also notice that, as $n$ gets larger, $x^{2n}$ have increasingly more weight on large-$x$ region.

To observe this, we first collect the exact result of $R_1(\lambda)$ and $T_2(\lambda,\lambda')$ without any approximation on $N$:
\begin{align}
&R_1(\lambda)=\sqrt{2N}^{1}K_{N}\left(\sqrt{2N}\lambda,\sqrt{2N}\lambda\right)\\
&T_2(\lambda,\lambda')=\sqrt{2N}^{2}\left[K_{N}\left(\sqrt{2N}\lambda,\sqrt{2N}\lambda'\right)\right]^2
\end{align}
with the kernel function given in terms of normalized wave-function of harmonic oscillator $\varphi_{j}(x)=(2^jj!\sqrt{\pi})^{-1/2}\exp(-x^2/2)H_j(x)$:
\begin{equation}
K_N(x,y)=\sum\limits_{j=0}^{N-1}\varphi_j(x)\varphi_j(y)
\end{equation}
\begin{figure}[t]
    \centering
    \includegraphics[width=0.49\textwidth]{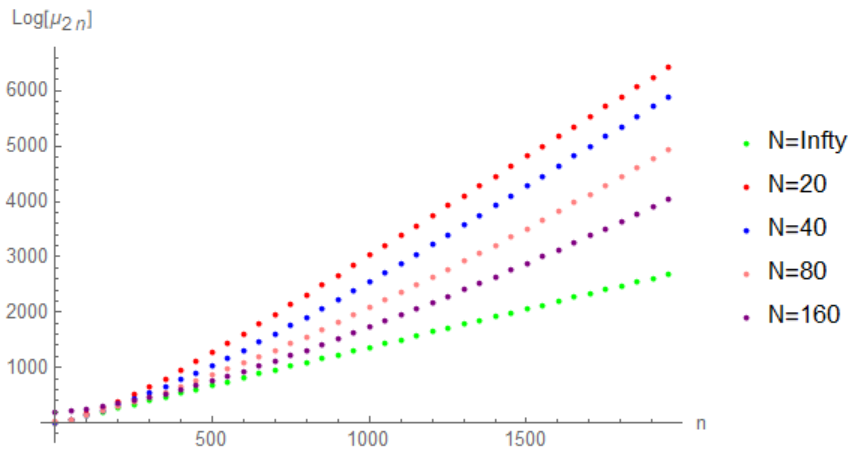}
    \includegraphics[width=0.49\textwidth]{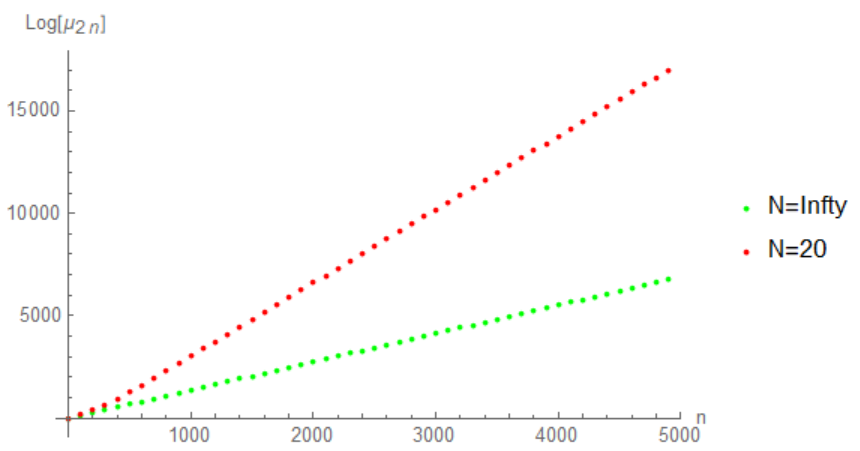}
    \caption{\textbf{Left:} $\mu_{2n}$-dependence on $n$, calculated at full $N$-dependence, from numerical integrating first line of equation~(\ref{eq:2}). \textbf{Right: }Same calculation as in the left panel, but to a larger scale.}
    \label{fig:finite N}
\end{figure}

Then from the exact expression of $\mu_{2n}$:
\begin{equation}
    \mu_{2n}=N^{-2}\int d\lambda a\lambda' (\lambda^2)(\lambda-\lambda')^{2n}\left[R_1(\lambda)R_1(\lambda')-T_2(\lambda,\lambda')\right],
    \label{eq:2}
\end{equation}
we obtain some numerical results shown in figure~\ref{fig:finite N}. We first indeed see that when $N$ becomes large, the result approaches infinite $N$ limit. In the right panel of figure~\ref{fig:finite N}, we see that even for finite $N$, the large $n$ behavior of $\mu_{2n}$ is still exponential in $n$. From equation~(\ref{eq:from b to mu}), a simple interpretation is that $\mu_{2n}$ would be exponential in the largest eigenvalue of $L$. 

We can also infer information of ${b_n}$ from the exponential growth.
Notice that this does not contradict with the fact that $b_n$ in large but finite $n$, namely $n\leq N^{2}$ would become zero, due to the termination of Krylov subspace. For example, for $N=20$'s case, we necessarily have $b_{400}=0$, and from equation~(\ref{eq:from mu to b})
we only need information of $\mu_{2n}$ at most up to $\mu_{400}$, so all the long exponential growth($\mu_{2n>400}$) in figure~\ref{fig:finite N} actually do not provide independent information, they only provide information on $b_{\max}$:
\begin{equation}
\max\limits_{0\leq n\leq N^{2}}[b_n]\sim\lim\limits_{n\rightarrow+\infty}\left[\frac{\log(\mu_{2n})}{2n}\right]
\label{eq:limit}
\end{equation}
So, the decrease and termination~\cite{Rabinovici2021} of $b_{N^2}=0$ actually hide in the subleading term on the exponential. 

The argument of equation~(\ref{eq:limit}) is in the following. For large $n$, $\mu_{2n}$'s dependence on ${b_n}$ is actually like a path integral:
\begin{align}
&\mu_{2n}=\sum\limits_{\text{path}}e^{S(\text{path})},\ \text{path}=(i_1,i_2,...,i_{2n}),\ \text{with}\ i_1=i_{2n}=0\\
&S(\text{path})=\sum\limits_{k=1}^{2n}\log b_{i_k}\approx 2n\oint dx\ \log(b(x))=2n\int_{0}^{1} dt\ \log(b(x(t))),\ \text{where}\ t=k/2n\\
&\Longrightarrow e^{\log(\mu_{2n})}\approx\int\mathcal{D}[x(t)]\ \exp\left[2n\int_{0}^{1} dt \log(b(x(t)))\right]\bigg|_{x(0)=x(1)=0,\ \dot{x}(t)<\epsilon}
\end{align}
For large $n$, this path integral is dominated by its saddle point, i.e., the particular path with the largest action. For finite $N$, $b_n$ would first grow and then decrease to zero~\cite{Parker2019}, meaning that there is necessarily a $b_{\text{max}}$, therefore, the saddle path would be `wandering around the site with $b_{\text{max}}$', which gives the argument of equation~(\ref{eq:limit}).

In the limit $N\rightarrow\infty$, the decreasing slope of $b_n$ would go to zero as $N^{-2}$~\cite{Parker2019}, then there are many paths with near the same action, and the saddle point approximation may not be valid.

In conclusion, although $b_n$ and $\mu_{2n}$ have an exact one-to-one mapping rule, this mapping is complicated. The large-scale structure of $b_n$ is hidden in the subleading terms and early-$n$-behaviour of $\mu_{2n}$, which is relatively hard to extract theoretically. 

\paragraph{Averaged Krylov complexity and Spectrum Form Factor}

As an aside, an interesting observation is that $\mu_{2n}$ in RMT is actually the moment of the superoperator. At infinite temperature, from equation~(\ref{eq:mu}) we observe:
\begin{equation}
\mu_{2n}=N^{-2}\sum\limits_{i,j}(\lambda_i-\lambda_j)^{2n}=N^{-2}\tr(\mathcal{L}^{2n})=N^{-2}(-1)^{n}\left(\frac{\partial}{\partial t}\right)^{2n}\tr(e^{i\mathcal{L}t})\bigg|_{t=0}
\end{equation}
which means that it is related to the higher derivative of SSF (Spectrum Form Factor) or analytically continued partition function. For finite temperature, from equation~(\ref{eq:mu, finite temperature}) we obtain:
\begin{equation}
\begin{aligned}
\mu_{2n}&=(-1)^n\left(\frac{\partial}{\partial t}\right)^{2n}\left[\sum e^{-\beta(\lambda_i+\lambda_j)/2+it(\lambda_i-\lambda_j)}\right]\bigg|_{t=0}\\
&=(-1)^n\left(\frac{\partial}{\partial t}\right)^{2n}\big|Z(\beta/2,t)\big|^2 \bigg|_{t=0}
\end{aligned}
\end{equation}
This observation also shows that the following quantity contains the same spectral information in random matrix theory:
\begin{equation}
\mathcal{K}(t)\longleftrightarrow\{b_n\}\longleftrightarrow\{\mu_{2n}\}\longleftrightarrow Z(t)
\end{equation}
Therefore, we may expect that the \textit{decay-dip-ramp-plateau} structure~\cite{Cotler2017} in SSF$(t)$ may translate correspondingly to some behavior in time dependence of complexity $\mathcal{K}(t)$. A possible obstacle is that the discontinuity of $Z(t)$ at dip-time and plateau-time, which is highly non-perturbative physics, would be hard to extract from its derivatives at $t=0$, which is perturbative. 

\section{Conclusions}
\label{sec:conclusion}
This study delves into the realm of Random Matrix Theory (RMT) to explore Krylov complexity $\K(t)$. In the large $N$ limit at infinite temperature, we analytically demonstrate the saturation of the Lanczos coefficient ${b_n}$ to a constant plateau $\lim\limits_{n\rightarrow\infty}b_n=b$, resulting in a linear increase in complexity $\K(t)\sim t$. Comparison of this plateau value $b$ with a diverse set of chaotic local quantum systems suggests its bounding effect on ${b_n}$ in such systems. Hence, we conjecture that, after scrambling time, the linear growth rate of Krylov complexity in chaotic local systems cannot exceed that in RMT.

At low temperature, we analytically establish that $b_n$ initially undergoes linear growth with $n$, reaching a plateau characterized by the renowned chaos bound. Upon reaching this common plateau $b$, $b_n$ stabilizes, indicating $\K(t)\sim e^{2\pi t/\beta}$ before the scrambling time $t_*\sim O(\beta\log\beta)$. Subsequently, a linear growth phase ensues, mirroring the behavior at infinite temperature. 

We conclude by addressing the influence of finite $N$ corrections, and the relation between K-complexity and spectrum form factor. Our results confirmed and strengthened previous understanding of RMT as a fast scrambler in the view of complexity growth. 

\acknowledgments
I would thank Chang Liu and Hui Zhai for previous collaboration on related topics. I would also thank Yiming Chen, Yingfei Gu, Chang Liu, Xiao-Liang Qi, Hui Zhai, Pengfei Zhang, Tian-Gang Zhou, and Yi-Neng Zhou for helpful discussion. I would especially thank professor Hui Zhai, not only for leading me to the topic of Krylov complexity but also for his encouragement and support throughout this project. The majority of this work was finished when I was studying at Institute for Adanced Study, Tsinghua University, and I would like to thank for their hospitality.

\bibliographystyle{JHEP}
\bibliography{biblio}
\end{document}